\documentstyle[prd,aps,epsfig,twocolumn]{revtex}
\begin{document}
\title{Dispersion Quark Model of Heavy Meson Decays}
\author{Dmitri Melikhov
\thanks{Talk given at the IVth International Workshop on Progess in Heavy Quark
Physics, Rostock, September 20-22, 1997}}
\address{Nuclear Physics Institute, Moscow State University, Moscow, Russia}
\maketitle
\widetext
\begin{abstract} 
We discuss applications of the dispersion quark model to exclusive 
semileptonic decays of heavy mesons. The transition form factors in this 
relativistic formulation of the quark model are given by double 
spectral representations through the wave functions of the initial 
and final mesons both in the scattering and the decay regions. 
An important feature of the model is that the form factors have 
the correct behavior in accordance with QCD in the case of meson decays 
induced by heavy parent quark transition both to heavy and light daughter quarks. 
We demonstrate that the dispersion quark model gives results compatible with the lattice calculations 
at large $q^2$ and thus provides reliable form factors in the whole decay region. 
\end{abstract}
\narrowtext
\vspace{10mm}
 
Theoretical description of hadronic amplitudes of the quark currents is one of
the key problems of particle physics as such amplitudes provide a bridge between QCD formulated in the
language of quarks and gluons and observable phenomena which deal with hadrons. In particular, the knowledge 
of such amplitudes is necessary for extraction of the parameters of the quark--mixing matrix in the
Standard Model from the experiments on weak heavy hadron decays. The difficulty in such calculations lie
in the fact that hadron formation occurs at large distances where perturbative QCD methods are not
applicable and nonperturbative consideration is necessary.

Various theoretical frameworks more or less directly related to QCD have been
applied to the description of meson transition form factors: among them lattice QCD \cite{lat}, QCD sum rules 
\cite{braun}, and constituent quark models \cite{alain}. 

Lattice QCD simulations is the most direct QCD based nonperturbative approach
and thus should provide most reliable results. 
Although it is not possible to place $b$ quark directly on the lattice, a constrained extrapolation in heavy quark mass 
\cite{lat} allows one to reliably determine the form factors for $B$ meson decay. A more serious problem is that 
lattice calulations do not provide the form factors in the whole accessible kinematical region: the daughter light quark 
produced in the $b$ decay cannot be allowed to move fast enough on the lattice and one finds himself constrained in the 
region of not very large recoils. For obtaining form factors in the whole kinematical decay region one can 
use the extrapolation procedures based on some parametrizations of the form factors. For instance, in \cite{lat} a 
simple lattice-constrained parametrization based on the constituent quark picture \cite{stech} and pole dominance is 
proposed. 
Anyway, a reliable knowledge of form factors in some region is already a substantial step forward which  
provides firm constrains on results of other approaches. 

QCD sum rules give in a sense a complementary information on the form factors as they allow one to determine 

$${}$$
\vspace{1.7cm}

\noindent the form factors at not very large momentum transfers \cite{braun}.  
However in practice various versions of QCD sum 
rules give rather uncertain predictions strongly
dependent on the technical subtleties of the particular version. 
A recent analysis \cite{ballbraun} disregards the three--point sum rules in favour of the light--cone sum rules. 
On the other hand, the light--cone sum rules involve more phenomenological inputs and the results 
turn out to be sensitive to a particular distribution amplitude of the light meson 
used in the evaluation of the sum rule and a model of subtracting the continuum (cf. \cite{aliev} and
\cite{damir}).  

Constituent quark models have proved to be a fruitful 
phenomenological method for the description of heavy meson transitions 
(a detailed review of the quark model is given in the talk by A. Le Yaouanc \cite{alain}). 
An attractive feature of the approaches based on the concept of constituent quarks is that these approaches
suggest a physical picture of the process. The physical picture of the constituent quark model is based on the
following phenomena responsible for the soft physics: the chiral symmetry breaking in the soft
region which yields appearance of constituent quarks and gluons; then,  
a strong peaking of the soft (nonperturbative) wave functions of hadrons with a width of order of the confinement scale; 
and, finally, the relevance of the contribution of only
Fock state components with a minimal number of the constituents (i.e. $q\bar q$ component in mesons and $qqq$
component in baryons). The latter assumption is justified by the successes of the constituent quark model
of hadron spectrum. We would like to illustrate that basically these features of the soft physics yield 
quite reasonable quantitative results for the meson transition form factors. 

An adequate framework which implements the abovementioned features of the nonperturbative physics 
is the dispersion formulation of the quark model of meson decays \cite{m1} which is based on taking into account
only two-particle $\bar qq$ intermediate states in the amplitudes of the $\bar q_s q\to \bar q_s q'$ 
transition through a bilinear quark current $\bar q' O q$. 

The transition form factors of the meson $M_1$ to the meson $M_2$ 
both in the scattering and the decay region are given by the double relativistic
spectral representations through the soft wave functions of the initial and final mesons, $G_1(s_1)$ and 
$G_2(s_2)$, respectively 
\begin{eqnarray}
\label{dr}
f(q^2)&=&\int \frac{ds_1\;G_1(s_1)}{s_1-M_1^2}\frac{ds_1\;G_2(s_2)}{s_2-M_2^2} \\
&\times&\left[\tilde f_D(s_1,s_2,q^2)+\tilde f_{\rm sub}(s_1,M_1^2,s_2,M_2^2,q^2) \right], \nonumber
\end{eqnarray}
where $s_1$ ($s_2$) is the invariant mass of the initial (final) $\bar qq$ pair. The double spectral 
density of the representation (\ref{dr}) consists of the two parts: the unsubtracted part $\tilde f_D$ 
and the subtraction part $\tilde f{\rm sub}$. 

The unsubtracted part $\tilde f_D(s_1,s_2,q^2)$ is calculated unambiguously from the double cut triangle 
Feynman graph with a relevant spinorial structure of the vertices corresponding to the initial and final 
mesons and to the quark transition current \cite{m1}. 

The subtraction part $\tilde f_{\rm sub}(s_1,M_1^2,s_2,M_2^2,q^2)$ contains the factors $s_1-M_1^2$ or $s_2-M_2^2$
and accounts for an ambiguity connected with possible subtractions in the dispersion representations 
inherent to any dispersion approach. To decide on the necessity and the structure of the 
subtraction terms we need some additional arguments. 

Notice that once we are working within an approach not directly deduced from QCD it is important to 
keep essential features of the underlying fundamental theory 
in the model. Thus matching the results obtained within the quark model to
rigorous QCD results might be helpful for bringing more realistic features to the model. 

QCD gives rigorous predictions for the structure of the form factors in the case of the meson transition induced by the
heavy quark transition. And it turns out possible to use matching the form factors of the dispersion quark model to 
form factors in QCD in the heavy quark limit for determining the subtraction terms in the spectral 
representations (\ref{dr}). 

Namely, we consider the case of a pseudoscalar heavy meson decay into final pseudoscalar and vector mesons induced 
by the heavy-to-heavy and heavy-to-light quark transitions through vector, axial-vector and tensor currents. 
We perform the expansion of the quark model form factors in inverse powers of the heavy quark mass 
$m_Q$ and match this expansion to the heavy quark expansion in QCD in leading and next-to-leading orders \cite{iwhh}. 
This matching provides constrains on the structure of the subtraction terms. 
Additional constraints allowing us to fix the subtraction terms are obtained by requiring the form factors of vector, 
axial-vector and tensor currents to obey the general relations found in \cite{iwhl} in the case of heavy-to-light 
transitions.  

It is important that the only property of the soft wave function of a heavy meson necessary for obtaining the correct
$1/m_Q$ expansion of the transition form factors is a strong peaking of this soft wave function in the relative momenta
of the constituent quarks in the region of order $\Lambda_{QCD}$. No other constraints on the soft wave function are
required. 

Finally, we arrive at the double spectral representations of the form factors with fixed subtraction terms. 
To obtain numerical predictions for the form factors we must specify the parameters of the quark model, 
i.e. constituent quark masses and the meson wave functions. 
To this end we take as an example the parameters of the ISGW2 model \cite{isgw2}. 
The form factors of the semileptonic decay of $B$ and $D$ mesons calculated within dispersion quark model 
adopting the ISGW2 quark masses and wave functions provide reasonable description of all experimental data 
\cite{m2}. Fig.1 illustrates the form factor $f_+^{B\to\pi}$ vs the lattice results, and Fig. 2 presents 
form factors for the transition $B\to K^*$. 

\begin{figure}
\centerline\mbox{\epsfig{file=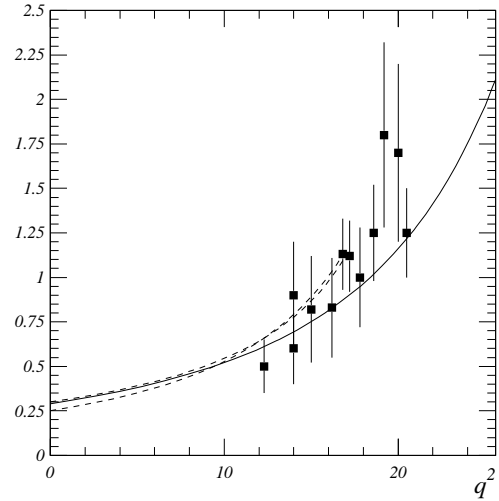,width=7.4cm}}
\caption{The form factor $f_+^{B\to\pi}.$ QM result-solid, 
LCSR \protect\cite{braun} - dashed,lattice points from \protect\cite{lat}.}
\end{figure}

\begin{figure}
\centerline\mbox{\epsfig{file=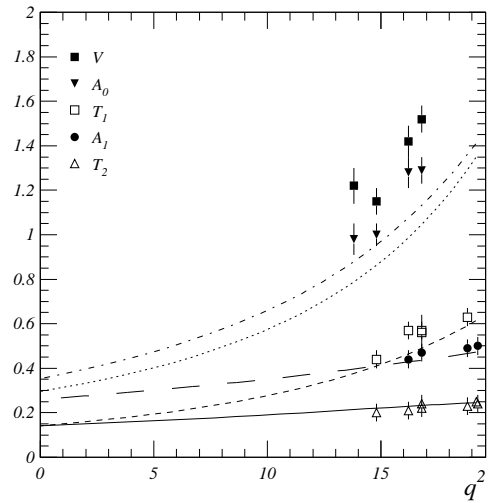,width=7.4cm}}
\caption{$B\to K^*$ form factors. QM results: solid-$T_2$, long-dashed-$A_1$,
dashed-$T_1$, dash-dotted-$A_0$, dotted-$V$. Lattice data from \protect\cite{lat}.}
\end{figure}

One can observe a reasonable agreement with the lattice results. 
For the value $T_2(0)=T_1(0)$ relevant for the $B\to K^*\gamma$ transition we find 
$$
T_2(0)=0.14. 
$$
which agrees well with a recent light-cone sum rule estimate \cite{damir}. 
 
It should be understood that the reported form factors are obtained with a very simple exponential ansatz for 
soft wave functions, which can be expected to describe only the effects of the confinement scale. 
Figs. 1 and 2 show that even this rather crude approximation works well and suggest that 
the bulk of the form factor behavior can be explained by the chiral symmetry breaking in the soft region and 
a strong peaking of the meson wave functions with a width of order of the confinement scale \cite{stech}. 

For obtaining more reliable predictions for the form factors one needs  
realistic meson wave functions and constituent quark masses relevant for the meson decays. These can be found by 
applying a combined fit based on the representations (\ref{dr}) to all available lattice points on meson transition 
form factors at large $q^2$ considering the wave functions and constituent quark masses as variational parameters. 
In this case the spectral representations of the dispersion
quark model may be treated as some generalized parametrizations of the transition form factors based on the 
constituent quark picture and obeying the rigorous QCD results in the limit of heavy-to-heavy and heavy-to-light
meson transitions. 

The work reported here was performed in part in collaboration with 
N. Nikitin and S. Simula and supported by the RFBR under grant 96-02-18121a. 
I take pleasure in thanking D. Becirevic, M. Beyer, V. Braun and A. Le Yaouanc 
for helpful discussions and the Organizers for the invitation to this pleasant and 
stimulating meeting. 

\end{document}